\documentclass[pdflatex,sn-mathphys]{sn-jnl}

\jyear{2021}%

\theoremstyle{thmstyleone}%
\theoremstyle{thmstyletwo}%

\theoremstyle{thmstylethree}%
\usepackage{mathrsfs,amsmath,amsthm}
\usepackage{booktabs}
\usepackage{longtable}
\usepackage{graphicx}
\usepackage{graphics}
\usepackage{hyperref}
\usepackage{braket}

\begin{document}

\title[The Quantum Internet : A Hardware Review]{The Quantum Internet : A Hardware Review}


\author*[1]{\fnm{Rohit} \sur{K. Ramakrishnan}}\email{rohithkr@iisc.ac.in}

\author[1]{\fnm{Aravinth} \sur{Balaji Ravichandran}}\email{aravinthb@iisc.ac.in}

\author[2]{\fnm{Ishwar} \sur{Kaushik}}\email{ishwarskaushik@gmail.com}

\author[3]{\fnm{Gopalkrishna} \sur{Hegde}}\email{gopalkrishna@iisc.ac.in}

\author[1]{\fnm{Srinivas} \sur{Talabattula}}\email{tsrinu@iisc.ac.in}

\author[4]{\fnm{Peter} \sur{P. Rohde}}\email{peter.rohde@uts.edu.au}

\affil*[1]{\orgdiv{Department of Electrical Communication Engineering}, \orgname{Indian Institute of Science}, \orgaddress{\city{Bengaluru}, \postcode{560012}, \state{KA}, \country{India}}}

\affil[2]{\orgdiv{Department of Electronics and Communication Engineering}, \orgname{National Institute of Technology}, \orgaddress{\city{Trichy}, \postcode{620015}, \state{TN}, \country{India}}}

\affil[3]{\orgdiv{Centre for BioSystems Science and Engineering}, \orgname{Indian Institute of Science}, \orgaddress{\city{Bengaluru}, \postcode{560012}, \state{KA}, \country{India}}}

\affil[4]{\orgdiv{Centre for Quantum Software and Information}, \orgname{University of Technology Sydney}, \orgaddress{\city{15 Broadway Ultimo}, \postcode{2007}, \state{NSW}, \country{Australia}}}


\abstract{In the century following its discovery, applications for quantum physics are opening a new world of technological possibilities. With the current decade witnessing quantum supremacy, quantum technologies are already starting to change the ways information is generated, transmitted, stored and processed. The next major milestone in quantum technology is already rapidly emerging -- the quantum internet. Since light is the most logical candidate for quantum communication, quantum photonics is a critical enabling technology. This paper reviews the hardware aspects of the quantum internet, mainly from a photonics perspective. Though a plethora of quantum technologies and devices have emerged in recent years, we are more focused on devices or components that may enable the quantum internet. Our approach is primarily qualitative, providing a broad overview of the necessary technologies for a large-scale quantum internet.}

\keywords{Quantum internet, Quantum computing, quantum communication, quantum cryptography, quantum photonics}



\maketitle

\tableofcontents

\section{Introduction}\label{sec1}

\textit{Quantum computing} has become a major buzzword of the current decade, and there is broad consensus that it will be one of the most critical future technologies. However, many scientists considered it unachievable until in recent years preliminary demonstrations proved otherwise. In the last decade quantum computing has finally escaped the realm of pure academic interest, with major industry players entering the quantum computing race and major developments rapidly taking place. The race for quantum supremacy has already been won in the context of NISQ (noisy intermediate scale quantum) devices, with the next major milestone being scalable, universal quantum computing, one that is likely to be achieved in the coming decade. This milestone will have far-reaching implications in technology, business, materials research, medicine, and everyday life via cloud-based applications.

The internet has arguably been the most transformative technology of the last two decades. With the advent of quantum technology, many possibilities are also opening in this field. A new form of the internet, the \textit{quantum internet}, is expected to emerge from the inception of new quantum technologies. Though it is still too early to predict the technologies that make this physically possible, photonics is foreseen as a core technology that will form the backbone of quantum internet technologies. Optics and photonics are already at the heart of many current communications technologies \cite{bib1}. Advances in this field lead us to conclude that photonics will also dominate the quantum internet's future. This paper provides an overview of quantum internet technology from an experimental photonics perspective. 

\section{Quantum computing}\label{sec2}

A quantum computer is a device that harnesses quantum mechanical phenomena to process information. The fundamental unit of a quantum computer is the quantum bit or qubit, which is the counterpart of bit in classical computers. Information processing in quantum computers uses a variety of quantum phenomena like quantum superposition and quantum entanglement, which give them the computational advantage over their classical counterparts.

Among the different approaches towards quantum computing, the two approaches that dominate the current technological scenario are circuit-based quantum computing and measurement-based quantum computing.

\subsection{Circuit-based quantum computing}\label{subsec2}

The most common model for quantum computing is the circuit model. This model gains wide acceptance due to its similarity with the classical computational models. The circuit models start with the input state, and the quantum circuits, which consist of quantum gates, manipulate the input states. The structure of the circuit and the gates are decided by the unitary transformation \(U\) of the circuit. The entanglement is created using the quantum gates, and the output is read out at the end of the computation process \cite{bib2}.

\subsection{Measurement-based quantum computing}\label{subsec3}

Measurement based quantum computing is a model of quantum computation proposed by Briegel et al. \cite{bib3}. They proposed a one-way quantum computer that uses cluster (or graph) states as resource states and showed that local measurements on individual qubits could realize any quantum circuit. This was further developed by many research groups, which led to a new computational approach, with entanglement as a resource and local measurements on qubits driving forward the computations. Adapting the measurement axes for future measurements, the randomness associated with the measurements can be handled, thus making the computation deterministic. Later works expanded the measurement based approach to quantification of entanglement for the scheme and to using cluster states and computational phases of matter and blind quantum computation, among others. It is regarded as a resource-efficient and alternative approach for the implementation of linear optical quantum computation put forward by Knill, Laflamme \& Milburn \cite{bib4}. There have been experimental successes in the implementation of cluster states and other resource states in various physical systems and measurement-based approach is a promising alternative to the circuit-based model of quantum computation \cite{bib5}.

Mathematically, a cluster (or graph) state is defined via the application of controlled-phase (CZ) gates to create edges on a graph between single qubits inititalised into the $\vert+\rangle$ state. Taking a graph defined by a set of edges and vertices $G=(V,E)$, a cluster state is given by,
\begin{equation}
    \vert\psi\rangle = \prod_{e\in E} \hat{\text{CZ}}_e \cdot \bigotimes_{v\in V}\vert +\rangle_v,
\end{equation}
where,
\begin{equation}
    \vert +\rangle = \frac{1}{\sqrt{2}}(\vert 0\rangle + \vert 1\rangle).
\end{equation}

Given a graph of appropriate topology, an arbitrary quantum computation may subsequently be implemented using a sequence of single-qubit measurements upon this state, where the graph structure and choice of measurement bases defines the implemented algorithm. The attractive feature of this alternate representation for quantum computation is that all entangling operations are shifted to a state preparation stage, after which execution involves only the most trivial of quantum operations.

Cluster states exhibit many useful properties that ease their construction, owing to the commutativity of CZ gates. Most notably, graphs can be partitioned, segments prepared independently in parallel, and then fused together. This is especially useful when dealing with non-deterministic entangling operations, as it overcomes the issue that all entangling gates must simultaneously succeed, which would have exponentially low success probability.

\section{Towards the quantum internet}\label{sec3}

The emergence of quantum computers has invariably led to the possibility of a quantum internet. More than just a network of quantum computers, it will be a network harnessing the laws of quantum physics and allowing quantum devices to exchange information. i.e., a quantum internet makes it possible to send qubits between multiple devices that are physically separated \cite{bib6}. This opens a new world of possibilities with unparalleled advantages compared to the existing internet, the classical internet. 

The hardware building blocks for such a quantum internet will constitute sources for quantum information, quantum repeaters, quantum memories and detectors for quantum information. This will also include integrated photonic technologies as platforms for quantum technologies and quantum satellites as part of the quantum networks.

\section{Sources}\label{sec4}

Sources for quantum information are the first building block of any quantum network. This can be quantum computers connected to the network or any quantum device that generate quantum information. A quantum source is any device that can successfully generate qubits with the required fidelity and accuracy. Optical photons are the most common sources for communication and computing applications. 

\subsection{Single-photon sources}\label{subsec4}

Single photon sources are crucial for several applications in quantum technology, including quantum communication, quantum metrology, optical quantum computation, quantum sensing, biomedical imaging, etc. The main feature of a single photon source is that they give us the ability to isolate and control a single quantum system. While we can already isolate quantum states like electron spins to form qubits, photons representing flying qubits open many possibilities, including quantum key distribution (QKD) and linear optical quantum computing (LOQC).

The ideal single photon source is an on-demand single photon source which produces a single photon in a known identical single mode each time it is called for. This ideal is impossible to achieve in practicality, but efforts are being made to achieve better and better approximations of the ideal.

Isolated single quantum systems that emit a single photon through processes like a radiative atomic transition are called deterministic sources. Unlike sources that produce single photons at arbitrary times, deterministic sources produce photons at specified times. This is an important distinction, although it can be assuaged through reconfigurable delay lines, mode-locked laser pumping, or other necessarily periodic storage cavities. Heralded sources, which involve the detection of one of a pair of photons, are called probabilistic sources. In principle, the time of arrival of the heralding signal is unknown. We know that a  single photon state has been produced only when a heralding signal is received. Probabilistic sources can become more deterministic through multiplexing of sources \cite{bib7}.

Spontaneous Parametric Down Conversion (SPDC) is a second-order nonlinear process that uses a high energy pump down-converted into lower energy signal and idler photons. SPDC is spontaneous because it can be triggered by vacuum fields, leading to interesting states. One application is the squeezed vacuum state, which is also used in gravitational wave interferometry \cite{bib8}. So-called phase-matching conditions ensure that SPDC obeys momentum and energy conservation. Since both the signal and idler photons are generated simultaneously, detecting one of them heralds the presence of the other, giving rise to heralded photon sources.

Four-Wave Mixing (FWM) is a third-order nonlinear process. Two pump photons are used to produce a signal and an idler photon. FWM also obeys phase-matching conditions but, unlike SPDC, has only been demonstrated in confined structures like integrated-optics geometries incorporating waveguides. The brightness of a source is a metric that has multiple definitions. For photon pair sources, the brightness is usually defined as the number of photons generated per pump pulse, normalised to pump power or photon bandwidth. FWM based sources have shown a brightness of around 0.855MHz \cite{bib9}.

In both SPDC and FWM, both emitted photons are correlated in time and are emitted simultaneously. Despite being probabilistic, one photon can herald the other due to this correlated emission detection. However, the approach is limited by multiple pair generation and loss. If either the signal or idler photon is lost, heralding does not occur. If multiple pairs are created, each beam of a heralded source setup would contain at least two photons. This can result in the heralding of more than one photon and have lasting repercussions in setups like those of QKD, wherein the security would be compromised due to the loss of uniqueness of the quantum information.

The photon statistics of a single photon source, or equivalently the second-order correlation function, is an important metric to judge a good source. An ideal single photon source will emit only one photon at a time. A method of measuring this metric is using a Hanbury-Brown and Twiss (HBT) interferometer. HBT interferometers consist of a beam splitter to split the input over two paths. The HBT will record only one detection event for an ideal single photon source. The time difference between two successive detection events can be plotted to deduce the emission statistics of the source \cite{bib10}.

Simple on-demand sources for single photons include atom/ion sources, diamond Nitrogen-Vacancy centre sources, and quantum dots. Atom/ion transitions are a simple method of photon emission. Excited single atoms can emit single photons spontaneously. However, isolating a single atom is a challenge. Trapped-ion and trapped-atom based single photon sources are also being developed. They demonstrate poor detection efficiency but were functionally the first single photon sources developed.

Defects in the carbon crystal lattice of a diamond can lead to a single carbon atom being replaced by a single nitrogen atom. This leads to a lattice vacancy known as a nitrogen-vacancy (NV) centre. Nitrogen is not the only atom that can develop such vacancies. Other atoms invading the diamond lattice can result in various diamond vacancy centres, known as diamond colour centres. However, NV centres show the most potential and even show promise for room temperature on-demand single photon source applications\cite{bib11}.

\subsection{Semiconductor-based photon sources}\label{subsec5}

Semiconductors are extremely promising as single photon sources, and many materials produce effective sources. This section will briefly review many of these semiconductor-based sources. For single photon emission, quantum dots based on III-V semiconductors have attracted much attention in recent years \cite{bib12}. 

Many materials form functional and high-performance single photon emitters at cryogenic temperatures. They include InGaAs and AlGaAs, among others. Several materials are under investigation for room temperature-based applications. Tamariz et al. \cite{bib13} have demonstrated quantum dot single photon sources operating at 300K using GaN as a base material. Fabricating such material under the constraint of integrability into photonic environments is a considerable challenge.

Quantum dots tend to be within the 2-10 nm range in size. On illumination by an optical pulse, electron-hole pairs are formed from the movement of electrons from the valence to the conduction band of the semiconductor quantum dot. These electron-hole pairs recombine for single photon emission. This pair production can occur multiple times leading to bi-exciton and multi-exciton states. Any of these transitions can be used for single photon emission through spectral filtering.

Quantum dots with both optical and electrical pumping have been realised. Electrically pumped sources have been achieved at room temperature. They show higher brightness on average, but optically pumped sources can show better anti-bunching, leading to higher confidence of accurate single photon production. Quantum dots integrated with cavities can show higher directionality in the emission of single photons. There is consistent progress in quantum dots as single photon emitters, and they are proving to be an effective deterministic source for applications like QKD \cite{bib14}.

Nanophotonic structures can enhance the emission of quantum dots and even help improve extraction efficiency. Within bulk semiconductors, quantum dots do not show high performance as single photon emitters. This is due to total internal reflection at the interface between semiconductor and air and a dearth of directionality in emission from quantum dots. Nanophotonic structures may improve extraction efficiency and spontaneous emission to the desired mode, inhibiting emission of other optical modes. In addition, nanophotonic structures offer on-chip integration possibilities \cite{bib15}.

Cavities keep light fields to confined volumes, allowing for resonant optical modes. Cavity structures can modify the density of optical states, allowing for spontaneous emission enhancement. With a given photonic structure, the Purcell factor gives the enhanced spontaneous emission rate ratio to the rate without the structure. So, increasing the Purcell factor increases the emission rate of single photons and the extraction efficiency of the source. The Purcell factor is related to the cavity quality factor. However, increasing the quality factor makes the frequency selectivity of the cavity increase as well, necessitating more accuracy in matching the quantum dot emission frequency with the cavity resonant mode.

For top-down single photon sources, distributed Bragg reflectors (DBRs) like planar and micropillar cavities are used for confining the light and forming a cavity. These consist of stacked alternating layers, which can be of a quarter-wavelength thickness and usually have different refractive indices. A semi-transparent DBR and a perfectly reflective DBR are combined to form the cavity, the semi-transparent part being used to extract the light from one side\cite{bib16}. 

Semiconductor defects also show promise for the development of single photon sources. The solid-state matrix of the host provides high stability, including at room temperature. This type of source can also be integrated into nanophotonic devices. Wide bandgap semiconductors like ZnO, SiC and diamond have also been used to demonstrate single photon sources successfully. Diamond and SiC especially have shown a wide range of optically active defects, of which some have shown characteristics of a single photon source \cite{bib17}.

Functional defects for single photon source fabrication are known as optically active defects. Optically active point defects are multi-level electronic systems with at least one radiative transition between levels. Defects can have complex level structures that include additional charge states by processes like ionisation of a nearby donor or have elementary two electronic level structures \cite{bib18, bib19}.

The emission rate of such systems can depend on factors like transition dipole moment or transition frequency. The environment can also play a role. After a system is excited, the probability of radiative decay is referred to as quantum efficiency. For a practical single photon source, we would like the emission rate to be as high as possible. This is because the emission rate defines the maximum pulse frequency that we can use to drive the source. For example, low pulse frequencies are unsuitable for QKD applications as that would lower the secure bit rate.

Some examples of colour centres that can result in defect-based single photon sources other than NV centres in diamond are divacancy (DV) of silicon carbide (SiC), diamond germanium vacancy, and silicon-vacancy in diamond, and rare-earth impurities in complex oxides. Recent studies show hexagonal Boron Nitride (h-BN) as a promising candidate for quantum point defect photonics applications \cite{bib20}. h-BN can show room-temperature single photon emission and displays a very high brightness \cite{bib21}.

Miniaturising quantum technologies to the chip level is critical as the field moves forward. This miniaturisation will improve cost, energy requirement, overall footprint, and reliability. On-chip photon sources tend to be based on nonlinear processes. Such integrated sources can be based on waveguides, or cavities \cite{bib22, bib24, bib25}.

A variety of methods can achieve such integrated sources, one being coupling a quantum dot to planar photonic crystal waveguides \cite{bib23}. Short optical pulses excite the quantum dot in this structure, causing a deterministic preparation of excitement in the quantum dot. The emitted photons are then channelled into a photonic crystal waveguide which is carefully designed such that the embedded quantum dot emits with a coupling efficiency of near unity. These collected photons can then be routed on-chip to tailored gratings where they can be outcoupled to optical fibres.

Nonlinear optics-based techniques also provide an intrinsic disadvantage to photon emission. These sources tend to be governed by statistical distributions such as Poissonian or thermal statistics, limiting the probability of single photon generation to as low as \textless25\%. Heralding can be a semi-effective workaround, allowing the single photon emission probability to increase without sacrificing the source quality, which is an effect of processes like multiplexing. Programmed multiplexers using FPGAs are also ways to improve the emission probability and generate better photon source circuits \cite{bib26}.

\subsection{Entangled photon sources}\label{subsec6}

Entangled photon pair sources, particularly polarisation entangled photon sources, are a vital resource for generating non-classical states of light. Typically, photon pairs created by  SPDC are entangled by the coherent superposition of two orthogonal SPDC processes in an interferometer. Entanglement and entangled photons have been exploited in several areas of quantum optics research, including computation, metrology and cryptography. 

Mathematically, the SPDC process is defined via a Hamiltonian that coherently exchanges photons between the laser pump source (with creation operator $\hat{a}^\dag_p$) and the so-called \textit{signal} and \text{idler} output modes ($\hat{a}^\dag_s$ and $\hat{a}^\dag_i$),
\begin{align}
    \hat{H} = \mathscr{E}(\hat{a}_s\hat{a}_i\hat{a}_p^\dag + \hat{a}_s^\dag\hat{a}_i^\dag\hat{a}_p),
\end{align}
Applying this Hamiltonian to a coherent state pump, we obtain the signal and idler output state,
\begin{align}
    \vert\psi\rangle = \sqrt{1-\chi^2}\sum_{n=0}^\infty \chi^n \vert n\rangle_s\vert n\rangle_i,
\end{align}
where $\chi$ is related to the interaction strength $\mathscr{E}$, the interaction time, and the amplitude of the coherent pump $\vert\alpha\rangle$. This creates an entangled superposition of photon pairs on the photon-number basis. Suppose a number-resolving photo-detector implementing the measurement projector $\vert 1\rangle\langle 1\vert$ is applied to the signal mode. In that case, we can see that a single-photon state in the idler mode is prepared.

There are multiple variants of this SPDC process. We have not denoted the polarisation degree of freedom in the above equations. In type-I SPDC, the photons have the same polarisation, while in type-II SPDC, they have orthogonal polarisation. In the latter case, creating a coherent superposition of the process with orthogonal polarisation allows us to directly prepare polarisation-encoded Bell pairs of the form $(\vert H\rangle\vert V\rangle+\vert V\rangle \vert H\rangle)/\sqrt{2}$.

Entangled photon sources can be operated in two different ways, through continuous-wave pumping and a pulsed pump laser. The pulsed laser method has been the default state of the art method since 2007, when a paper was published by Ma et al. detailing a QKD scheme using such sources \cite{bib27}. Neumann et al. have shown QKD with continuous wave (CW) entangled photon pair sources \cite{bib28}. Beam displacer-based designs are also quite popular, a few given descriptions by Fiorentino et al. in 2008 \cite{bib29}. Birefringent crystal is cut so that the optical axis forms an angle with the direction of propagation of the optical beams to form a beam displacer for these designs.

In addition to polarisation, frequency is a degree of freedom commonly used in practical applications. Riazi et al. have shown an entangled pair shaping method in frequency and polarisation degrees of freedom by cascading entangled photon pair sources \cite{bib30}. The frequency-domain offers an exciting framework for producing high-dimensional entangled quantum states on a single photonic chip through techniques like spontaneous four-wave mixing (SFWM) \cite{bib31}. Photon pairs generated using SPDC are generally correlated in frequency, described by a Joint Spectral Amplitude (JSA). Techniques like quasi-phase-matching (QPM) can be used to shape the spectrum for applications like the generation of single photons from these pairs through detection to generate heralded photons. Dosseva et al. have shown a custom poling optimisation to filter out the correlations between the pairs spectrally \cite{bib32}.

Entangled photon pairs are a relatively accessible source of entangled quantum states for applications such as testing for local realism \cite{bib33}. Photons are also highly promising as quantum bits or qubits because photonic qubits can move at the speed of light, experience virtually no decoherence, and operate well at room and cryogenic temperatures. An exciting application for photonic qubits is in scalable quantum information processing, particularly the use of entangled photon pairs in the frequency domain \cite{bib34}. 

In traditional, more popularly used sources such as interferometer-based entangled photon pair sources, a phase discrepancy is possibly introduced by dispersion \cite{bib35}. A plausible solution could be to change the materials used for beam displacement in the interferometer to materials with similar dispersion, but this solution may not always be viable. Horn and Jennewein have shown a double displacement method to work around this drawback and produce a robust interferometer design for polarisation entangled photon pair sources \cite{bib36}.

Semiconductor Quantum Dots can be used as a source of polarisation entangled photon pairs \cite{bib37}. The process by which the entangled pair is generated is known as the biexciton-exciton cascade. The quantum dot is initially in the biexciton state, which then decays through recombining one of two electron-hole pairs. A single photon is emitted by this process, leaving the quantum dots in an exciton state, after which it decays once more through the emission of another single photon.
Quantum dots show an entanglement fidelity of up to 97.8\%.

Historically, the significance of quantum dots as emitters of entangled photons has been crippled by the number of various underlying effects that can degrade entanglement. However, many of these effects have been worked around in recent times. These effects include fine-structure splitting (FSS), valence band mixing, recapture processes, and exciton spin-flip process.
	
\subsection{Photon sources for satellites}\label{subsec7}

We require sources of single quantum states for transmission for long-distance communications like satellite-based quantum communication. Different QKD applications, such as Measurement Device Independent (MDI) QKD or Discrete Variable (DV) QKD, also require single photon sources. Phase randomised WCP sources are used in protocols like BB84, Weak Coherent Pulse (WCP), Coherent One-Way (COW), and Decoy state (DSP).

A laser pulse is attenuated to the degree where the average photon number per pulse is less than one, which simulates an ideal single photon source since it obeys Poissonian statistics. WCP sources have a small multi-photon emission probability, resulting in information leakage to eavesdroppers. However, the protocol ensures safety even with this drawback.

WCP sources have proven robust and straightforward for satellite applications. However, continuous variable sources and entanglement sources are proving to be more and more usable in recent times, opening a possibility for their use in satellite QKD applications. Some shortcomings of WCPs include side-channel information leakage in auxiliary degrees of freedom and imperfect overlap of spectra of pulses, among others. These can reduce the security of the protocol. WCPs still have the highest transmission rate, showing GHz performance for on-chip platforms.

Quantum dots integrated into 2D materials could be close to accurate single photon sources for satellites that do not require cryogenic temperatures to operate. Along with their high brightness metric, these sources have suitable characteristics for quantum communication links. 

Entanglement achieved through SPDC, waveguide sources, FWM, or any other method described earlier can also be helpful in satellite quantum communications. Entangled states can be used for dense coding wherein more than one bit of information is transmitted per photon or pulse. Quantum communication does not require entanglement, but several protocols rely on entanglement, which motivates to development of satellite-based entanglement distribution networks \cite{Sidhu2021}.

Finally, continuous variable (CV) protocols are also incorporated in satellite-based communications. They encode information in degrees of freedom, such as field quadrature. CV protocols use infinite-dimensional Hilbert spaces to show higher key rates as they could encode more information per pulse. CV sources generally base themselves on Gaussian-modulated coherent states, which are simple to generate and detect.

\section{Photo-detectors}\label{sec5}

Photodetectors are vital devices for quantum networks as they convert the incoming optical signals into electrical signals. Semiconductor photodetectors or photodiodes are the most popular photodetectors because of their faster detection speed, smaller size, and higher efficiency. 

Mathematically a generic photo-detector operating in the photon-number degree of freedom is modelled as a positive operator-valued measure (POVM),
\begin{equation}
\hat{\Pi}_m = \sum_{n=0}^\infty P(m\vert n) \vert n\rangle\langle n\vert,
\end{equation}
where $\hat{\Pi}_m$ is the measurement projector for the $m$-photon outcome event, and $P(m\vert n)$ is the conditional probability of measuring $m$ photons given $n$ incident photons. This can also be viewed in the quantum process formalism as,
\begin{equation}
\mathcal{E}_m(\hat\rho) = \sum_{n=0}^\infty P(m\vert n) \hat{E}_n\hat\rho\hat{E}_n^\dag
\end{equation}
where $\hat{E}_n=\vert n\rangle\langle n\vert$ are the Krauss operators, in this instance the $n$-photon projection operators.

Using this representation, the $P(m\vert n)$ matrix fully characterised the detector's operation in the photon-number degree of freedom. For a simple lossy detector with no dark-counts, we have,
\begin{equation}
    P(m\vert n) = \binom{m}{n}\eta^n(1-\eta)^{m-n},
\end{equation}
where $\eta$ is the detection efficiency. We can also represent `bucket` detectors, which only differentiate between no photons and some photons as,
\begin{eqnarray}
    \hat{\Pi}_{\text{off}} &=& \hat\Pi_0,\nonumber\\
    \hat{\Pi}_{\text{on}} &=& \sum_{m>0} \hat\Pi_m = \hat{I} - \hat{\Pi}_{\text{off}}.
\end{eqnarray}

\subsection{Multiplexed photo-detectors}\label{subsec8}

The class of photo-detectors which are experimentally challenging to realise are photon number resolving detectors. However, non-number-resolved detectors can be used to approximate number resolving detectors using multiplexing techniques. This is more complex in terms of experimental realisation. An $n$-photon state is evenly spread across many modes $m$ which are independently measured using non-number-resolved photo-detectors. We can conclude that more than a single photon has reached some detectors if $m< n$. Similarly, if $m>>n$ it becomes highly unlikely that more than a single photon reaches a given detector, in which case the sum of detection events closely approximates total photon number. Spatial and temporal domain multiplexing is possible but with essential experimental overheads \cite{HUI2020125}. 

Based on this principle, the probability that a mutiplexed detector based on non-number resolving detectors measures the same number of photons as are incident is given by,
\begin{equation}
    P(n_\text{meas}=n_\text{inc}) = \frac{\eta^n m!}{m^n(m-n)!},
\end{equation}
where $\eta$ is the detection efficiency. In the limit of a large number of multiplexed modes $m$ we observe ideal operation,
\begin{equation}
    \lim_{m\to\infty} P(n_\text{meas}=n_\text{inc}) = \eta^n,
\end{equation}
which gives $P(n_\text{meas}=n_\text{inc})=1$ for detectors with perfect efficiency.

\subsection{Homodyne detectors}\label{subsec9}

Homodyne detectors are used in continuous-variable systems rather than discrete variable systems (photon number states). In this method, the input beam and a reference beam (used as a phase reference) are made to interfere through a beamsplitter, and photo-detectors are used to detect both output modes, and the difference in photon count rates is taken. We can directly sample points in phase-space, thus constructing the Wigner function of the unknown state by sweeping through the amplitude and phase of the reference beam. The experimental challenge lies in preparing a coherent beam that is used as the reference.

\subsection{Practical photo-detectors}\label{subsec10}

Practical photodiodes can include several adjustments in their structure to enhance quantum efficiency. An example of this is the PIN diode which has an intrinsic layer between the P and N layers. An often-used detector is an Avalanche Photodiode (APD). APDs can introduce considerable photon amplification using avalanche gain, given that the bias voltage is high enough. APDs tend to show high amounts of dark current and are usable with moderate dark current at low temperatures. However, Jones et al. have recently demonstrated high-temperature APDs using AlInAsSb and GaSb \cite{Jones2020}.
	
An ideal photodiode should instantly translate each photon from the optical signal into an electron. This would imply that the photocurrent is linearly proportional to the power of the optical signal. Practically, however, not every incoming photon may produce an electron. This can be caused by inefficient photon absorption and carrier collection. The speed of photodetection may also be affected by transient carrier effect or RC parasitic behaviour. So, the responsivity of photodiodes depends on many factors like semiconductor bandgap structure, material quality, photonic device structure, etc. Practical photodiodes also generate noise that can degrade an optical communication system's signal-to-noise ratio (SNR). Some familiar noise sources include shot noise, thermal noise, and dark current noise. Out of these three, shot noise is the only form of noise that cannot be reduced. It is fundamentally associated with photodetection and hence introduces the fundamental limit to the performance of an optical system known as the quantum limit.

The photocurrent is proportional to the received optical signal power. Hence, photodiodes are square-law detectors. This can create additional problems such as mixing between frequency components of the optical field introducing beating noises in the electrical output. Another type of photodiode that is commonly used is a Photomultiplier Tube (PMT). PMTs consist of a photocathode followed by an electron multiplier. When a single photon causes the ejection of an electron from the photocathode, electric fields in the PMT accelerate it into another surface known as a diode. Several electrons are released upon collision with the diode, which is accelerated into collisions with another. The process is repeated, and a typical electron gain of $10^6$ is usually produced.

A better signal-to-noise ratio can be achieved by directly measuring the photocurrent from the output of the PMT, using a nonlinear process known as photon counting. Each photon incident on the PMT produces a discrete pulse at very low optical intensities. Current pulses of $\sim10^6$ electrons are counted. The sum of counts over a given time is proportional to the optical intensity. The probability of two or more photoelectrons being produced simultaneously is negligible at low count rates. Simultaneous counts cause nonlinearity between the count rate and optical intensity at higher optical intensity. PMTs have the lowest dark noise per unit detection area of any detector, and they have quantum efficiencies of $1-40\%$ in the ultraviolet and visible regions. The efficiency falls off rapidly in near-infrared regions.

A complete comparison between APDs and PMTs was drawn by Wernick et al., wherein they tabulated the gains, quantum efficiencies, max dark current, and other essential properties \cite{alma9924392413402466} The results can be summarised as follows. APDs show significantly less gain than PMTs ($\sim10^2$ vs $\sim10^6$)  and have much longer signal rise times but display much higher quantum efficiencies ($66\%$ vs $24\%$ at 400 nm). An APD array performs better in most fields than a PMT, the only drawbacks being the gain mentioned above and signal rise time. PMTs also tend to have variations in gain and linearity at higher counting rates, as shown by Yu et al.\cite{2021NIMPA100865433Y}. Other photodetector materials have been explored, such as In\textsubscript{2}Te\textsubscript{3}, ZnO, and GaN. These materials have shown low noise, high photosensitivity, and good stability. However, several new materials suffer from limited photocurrent and photoresponse speed.

Nanostructured materials have been developed to get around these limitations. Nanoscale elements like Quantum Dots (QDs), nanowires, and nanolayers can be used to fabricate high-performance detectors. They show impressive performance in various fields such as high sensitivity, high miniaturising potential, fast response, and multifunctionality. 2D materials show many favourable properties for photodetection. They are usually fabricated by more straightforward techniques than semiconductor films which use universal epitaxial techniques. 2D nanomaterials also show great potential in optoelectronic and electronic applications. For example, graphene shows high charge-carrier mobility, broadband absorption, and optical transparency. In addition to these features, graphene exhibits good mechanical flexibility. Because of the typical plasmonic structures of graphene, photocurrent is no longer limited to photon energies greater than the bandgap of a semiconductor. Additionally, it can be integrated into almost any waveguide material like SiN, Si, or AlN.

Schuler et al. put forth an ultrafast photodetector using graphene on a Si slot waveguide \cite{Schuler2016}. This photodetector showed a responsivity of 76 mA/W at a moderate drain-source bias of 300 mV and recorded a 3dB bandwidth of 65 GHz. However, realisation of broadband photodetectors is hindered by low absorption in visible and near IR regions exhibited by graphene. However, there are methods to work around this problem, such as the stretchable solution presented by Kang et al. \cite{Kang2016}. Zhuo et al. have demonstrated a graphene-based photodetector that integrates multiple fibers, including carbon nanotubes (CNTs), resulting in increased responsivity \cite{Zhuo2020}.

Transition Metal Dichalcogenides (TMDC) show an advantage over graphene in one key aspect- dark current. Graphene shows a high dark current due to its gapless nature. This has a significant impact on the sensitivity of photodetection. TMDCs have a natural gap and have captured attention due to that advantage over graphene. Several new photodetectors using TMDCs like MoS\textsubscript{2}, WS\textsubscript{2}, InSe, GaS, GaTe, WSe\textsubscript{2}, GaSe, and In\textsubscript{2}Se\textsubscript{3}. An example is the MoS\textsubscript{2} photodetector proposed by Mao et al. \cite{Mao2021}. Another recent layered material used for photodetection is Black Phosphorous (BP). BP has a tuneable direct bandgap covering the optical spectrum range from visible to NIR. BP also shows good electrical properties such as high hole mobility and good current saturation. It also has a unique set of in-plane anisotropic physical properties. These factors prove BP to be the most promising candidate for broadband photodetection. However, in BP-based photodetectors, certain factors such as photoresponsivity and response time are still unsatisfactory. Li et al.  have given an overview of BP-based photodetectors and some methods to overcome the challenges mentioned \cite{Li2018}. Ma et al. have proposed a method to achieve high responsivities with BP photodetectors utilising the slow light effect of photonic crystal waveguides \cite{Ma2020}. 

Another upcoming type of photodetector is perovskite-based photodetectors or PPDs (Perovskite Photodetectors). Perovskite semiconductor materials show superior electronic and optical properties and make for very promising photodetectors for various practical applications. In recent years, many techniques for film deposition and solution synthesis control the morphology and composition of perovskite materials have encouraged an increasing amount of attention towards PPDs. Organic Photodetectors (OPDs) are a promising approach with advantages such as low cost, tunability of detection wavelength, and compatibility with flexible and lightweight devices. Inorganic Photodetectors (IPDs) suffer from drawbacks like brittleness and complicated manufacturing. Spectrally selective IPDs require optical filters to be attached. OPDs can overcome these drawbacks. UV, visible, and NIR OPDs can be achieved by optimising the optical bandgap of the organic semiconductors or through particular device architecture instead of attaching optical filters \cite{Miao2019}.

For highly sensitive photodetectors, an External Quantum Efficiency (EQE) of greater than 100\% is desirable. OPDs are capable of Photomultiplication (PM), which displays this required threshold of EQE. These PM OPDs are used further to realise PM PPDs. Perovskite-organic hybrid photodetectors that display advantages of both perovskite and organic semiconductors show promising results such as the extension of detection region and unique wavelength detection. Filter free narrowband PM OPDs have been realised, which paves the way for filter-free narrowband PM PPDs with an EQE of 100\% \cite{Wang2017, Ren2021}.

\begin{table}[b]
    \centering
    \includegraphics[width=11cm,height=9cm,keepaspectratio]{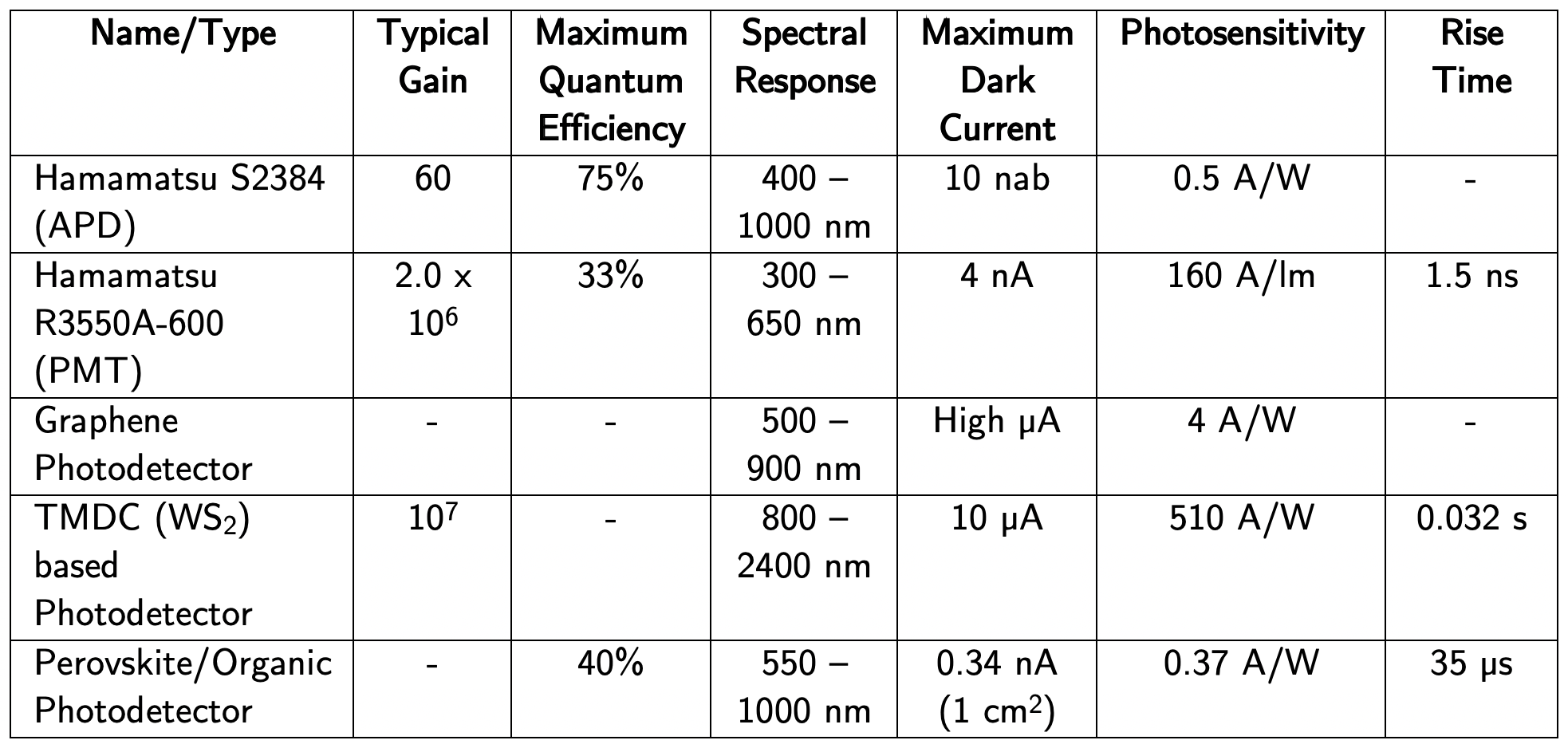}
    \caption{Different types of photodetectors and their characteristics}
    \label{Table1}
\end{table}

Frequency up-conversion photon detection schemes are techniques implemented to convert photons to a wavelength that can be detected efficiently by a single-photon detector. One such mechanism used is sum-frequency generation through a nonlinear optical crystal. This upconversion can be achieved with very high efficiency with sufficient pump power. Technical challenges include achieving the desired field strength, methods for which are summarised by Hadfield \cite{Hadfield2009}. However, stabilising the nonlinear crystal is quite tricky, and if waveguides are used in the scheme, they can display coupling losses. Another detection method is Visible Light Photon Counters (VLPCs). VLPCs offer high detection efficiency for single photons but only up to a wavelength of $1  \mu m$. VLPCs can resolve photon numbers and display a good timing resolution. They show a moderate number of dark counts, however. VLPCs are based on solid-state photomultipliers, which are devices that use As-doped silicon.

The latest trend in the field of quantum detection includes the superconducting nanowire single-photon detector (SNSPD). It is a quantum-limit superconducting optical detector based on the Cooper-pair breaking effect by a single photon. They exhibit higher detection efficiencies, lower dark count rates, higher counting rates, and lower timing jitter than their counterparts. SNSPDs have been extensively applied in quantum information processing, including quantum key distribution and optical quantum computation. SNSPD is based on the narrow, thin superconducting nanowire, current-biased close to its critical current, typically operating at around 1K. When a photon is absorbed in the detection region, its energy drives the wire normal, producing a voltage pulse. The SNSPD offers high timing resolution and detection of the photon flux at the shot-noise limit due to the speed of the detection process. The combined effect yields a detector that can measure the photon’s arrival time from a faint object or source. SNSPDs also offer single-photon detection capability beyond the wavelength detection range of silicon and InGaAs, to at least 5000 nm \cite{Marsili2013, You2020}.

In contrast to the SNSPD, the Superconducting transition-edge sensors (TESs) are photon-number resolving detectors with almost unity detection efficiency. They offer high-efficiency single photon detection with low dark counts and photon number resolving properties. They operate at low temperatures, and any temperature change can cause an abrupt change of resistance in the superconducting film, which is kept on the cusp of a superconducting transition. Incident photon absorption catalyses this temperature change causing a voltage-based detector to draw a current, amplified for photo-detection. At a fixed wavelength, the signal becomes proportional to the photon number. Room temperature black-body radiation can affect the dark count rates in such a detector \cite{Gerrits2016}. Assuaging this effect through filtering is possible, but this decreases the detector efficiency. Due to its energy resolving capability, the TES can also be used as a spectrally resolving single-photon detector with limited resolution \cite{Michael}.

A summary of commercially available detectors is presented in Table 1. \cite{Liu2014, Onur, Li2020}

\section{Quantum memories}\label{sec6}

Quantum memories are fundamental building blocks to any global-scale quantum internet, high-performance quantum networking and near-term quantum computers. Such a memory must be able to store qubit strings (or part of qubit strings) faithfully and release them on demand \cite{Lvovsky2009}. However, such storage need not be perfect. If the fidelity of the memory falls below a certain threshold, fault-tolerant schemes can be utilised for their successful operation \cite{Gyongyosi2020}. A generic quantum memory shall store a pure or mixed quantum state, that can be denoted by its density matrix $\rho$ and shall output a state $\hat\rho$ which is close to $\rho$ \cite{Liang2019}. The storage fidelity of a quantum memory can be mathematically defined for a specific state as  
\begin{equation}
    F(\rho)=Tr \sqrt{\sqrt{\hat\rho} \rho\sqrt{\hat\rho}}
\end{equation}

Therefore, there is a threshold for the worst-case fidelity for every quantum memory beyond which the fault-tolerant quantum error correction techniques are applied. The calculation of fidelity is done by characterisation of the quantum process, and it is a comparatively cumbersome process experimentally\cite{Lvovsky2009}. Another parameter used to characterise quantum memory is its efficiency $\eta$, defined as the ratio between the energies of the stored and retrieved signals. It can be evaluated experimentally, but it is still not a valid measure as it does not consider data contamination from the storage medium. The multi-mode capacity determines the number of optical modes that can be successfully stored in a quantum memory, dependent on the mechanism used to realise the memory. Another essential parameter for quantum memory is storage time. We could call the ratio between the storage time and the duration of the stored pulse the delay-bandwidth product as the figure of merit for most of the applications.

\subsection{Optical delay lines and cavities}\label{subsec11}

As an optical fibre, an optical delay line is the most straightforward approach to quantum memory. The fundamental limitation of this approach is that its storage time is fixed by the delay length, which may not be ideal for some of the on-demand applications. An alternate method is to store light in a high-Q cavity where it cycles back and forth between the reflecting boundaries. The signals can be sent in and out of the cavity by electro-optical or nonlinear optical methods or transfer the quantum state to passing atoms. However, the efficiency of such cavities is limited by the trade-off between short cycle time and long storage time \cite{Heshami2016}.

\subsection{Electromagnetically induced transparencies}\label{subsec12}

A nonlinear approach for quantum memory is Electromagnetically induced transparency (EIT). Atoms with discrete energy levels - two optical fields, a robust control field and a weak signal field, are utilised. The weak signal carries the quantum information while the strong control field steers the atomic system by coupling the excited level to their ground levels. In the absence of a control field, the signal field is partially or entirely absorbed by interacting with the resonant two-level system. When the frequency difference of the optical fields is close to the Raman transition between the ground states of the system (two-photon interference), the absorption is mainly reduced in the presence of the control signal \cite{PhysRevA.104.063704, Ma2017}.

The signal pulse is fed to the cell under EIT conditions when switched on the control field. The control field is then switched off while the spatially compressed pulse propagates inside the EIT cell, allowing the quantum information carried by the pulse to be stored as a collective excitation of the ground states. The control field is switched on again when the pulse needs to be retrieved,  and the pulse exits the cell \cite{Novikova2012}.

\subsection{Photon-echo quantum memory}\label{subsec14}

Photon-echo quantum memory is in principle like the EIT schemes. This method transfers the quantum state carried by an optical pulse to a collective atomic excitation. Unlike EIT, photon-echo quantum memory takes advantage of inhomogeneous broadening. 

Consider an ensemble of two-level atoms in the state $\vert\psi\rangle$ After absorption of a signal photon at $t = 0$, the state is 

\begin{align}
    \vert\psi\rangle = \sum_{j}\psi_{j}e^{-i\delta_jt}e^{ikz_j}\vert b_{1\ldots}a_{j\ldots}b_N\rangle
\end{align}

where $k$ denotes the wavevector of the signal field, $\delta_j$ is the transition detuning of atom $j$ with respect to the light carrier frequency. 

The atomic dipoles are initially phase-aligned with k, but this alignment rapidly decays since $\delta_j$ differs for each atom. All photon-echo quantum memory protocols use a procedure that rephases the atomic dipoles later, recreating collective atomic coherence. In other words, the phases of all atoms become equal at some point in time $t_e$, and this triggers re-emission of the absorbed signal \cite{Tittel2009, Ma20212}.

\subsection{Nano-Opto-Electro-Mechanical-Systems}\label{subsec15}
Nano-Opto-Electro-Mechanical-Systems (NOEMS) constitute an emerging technology area as they provide platforms to investigate nanophotonic materials' electronic and mechanical properties. Their optical, electrical, and mechanical degrees of freedom can be used to manipulate information carriers. NOEMS can be used to make high speed and low-power switches, high-efficiency microwave-optical conversion devices, and multiple quantum information processing devices through on-chip integration \cite{Xu2021}.

Advances in cavity quantum mechanics have led to experimental work on embedding high-Q mesoscopic mechanical oscillators into microwave and optical cavities. For microwave to optical conversions, the most effective method is the optomechanical method, with an efficiency of 47\%. In a quantum network, infrared or near-infrared photons are used as flying qubits to transmit information and microwave photons are used to manipulate information. NOEMS devices facilitate the bidirectional transfer between microwave and optical photons. On-chip NOEMS are compatible with superconducting circuits and provide an excellent candidate for quantum memories for storing information. The memory function is based on the geometry of the devices, which allows flexibility in materials and designs. Wallucks et al. recently demonstrated a quantum memory in telecom wavelength using an engineered high-Q mechanical resonant cavity and an optical interface over the full decay time of the mechanical mode \cite{Wallucks2020}. The quantum memory is prepared in a single-phonon state, directly usable for a Duan–Lukin–Cirac–Zoller-type (DLCZ) quantum repeater scheme. They showed that this state could be stored for ~2 ms without degradation and retrieved optically by a weak coherent probe to measure the coherence of the mechanical mode.

\subsection{Atomic frequency combs}\label{subsec16}

In this method, the distribution of atoms over the detuning  $\delta$ is expressed by a periodic comb-like structure with absorption lines separated by multiples of $\Delta$   (difference in energy levels). Repetitive rephasing occurs at times $2\pi/\Delta$, when the phases accumulated by atomic dipoles in different `teeth' differ by multiples of $2\pi$. When the comb structure is absent, the excited optical coherence can be temporarily transferred to other atomic levels to prevent re-emission after a fixed cycle and facilitate long-term storage and on-demand readout \cite{PhysRevA.99.052314}.

Overall, the quantum memories for quantum internet are still very much in the development phase in terms of research. It is too early to predict in terms of technology which mode of physical realisation will be the forerunner in the game.

\section{Quantum repeaters}\label{sec7}

Repeaters are a vital component in any long-distance communication system. Quantum repeaters are indispensable to transmitting quantum information over arbitrarily long distances in the quantum internet. Quantum repeaters are needed when there is no direct line of communication between a server and two clients, and high-quality entanglement must be shared between nodes. This can be done using bootstrapped entanglement swapping and purification protocols. Bell pairs are the most used entanglement, which serves as a resource for most quantum protocols. The links over shorter distances can easily be realised with much higher success probabilities, and they can be stitched together by entanglement swapping. This is how long-range communication links are generated.

Any quantum repeater network performs three main operations:
\begin{enumerate}
\item Entanglement distribution: creating entangled links between adjacent repeater nodes. 
\item Entanglement purification: improving the quality of entanglement between nodes.
\item Entanglement swapping: joining adjacent entangled links together to form longer distance links. 
\end{enumerate}

The first step in preparing a long-distance quantum communication link is the preparation of multiple entangled pairs between adjacent repeater nodes. The methods used for entanglement distribution are mainly classified into three.
\begin{itemize}
\item Photon emission from quantum memories in the repeater nodes, followed by which-path erasure
\item Absorption of entangled photons by quantum memories
\item Photon emission at one node and absorption at another 
\end{itemize}

The emission based methods are the most popular ones. Once we have enough entangled pairs, we move to the following purification process. Long distance quantum communication links require a certain amount of fidelity. The generated entangled states may lack the same due to the finite coherence time of quantum memories, operation errors of quantum gates and large spatial separation. The purification process involves the usage of probabilistic heralded error detection codes or error correction codes. Implementation of the error correction codes requires stringent conditions on the initial fidelity of entangled states and the quality of quantum gates.

The distribution and purification processes generate high-fidelity entangled states between adjacent repeater nodes. The next task is to extend the range of entangled states, which is done by entanglement swapping. Consider a Bell-pair held by two parties. The entanglement swapping is done by taking those Bell pairs and swapping the entanglement between them such that the entangled state is now shared between two parties. This process can be bootstrapped progressively, and the entanglement can be swapped over longer distances, yielding quantum repeater networks. 

The first generation of quantum repeaters was probabilistic by nature. Classical signalling was required between nodes to inform successes or failures. Thus, the performance of repeater networks was constrained by the probabilistic nature of the involved quantum operations and associated classical communication time. This created the need for similar operations deterministic in nature. 

The second generation quantum repeaters used error correction codes instead of the error detection codes used in early quantum repeaters. Typical protocols in early repeaters detect errors and discard the associated entangled pairs, while the error correction codes correct some of the errors, avoiding the need to discard the affected entangled states completely. However, the error correction schemes are deterministic by nature, and that demands high fidelity in the generation of entangled states initially between adjacent nodes. Still, the second generation repeaters had substantial improvement over the first generation repeaters in terms of time and resources, at the price of higher fidelity required initially. If a fidelity over $50\%$ was sufficient for first-generation schemes, the second-generation methods with error correction protocols had to more than $90\%$ . Though the performance was increased in the second generation schemes, the communication time between adjacent repeater nodes was limited to determine the success of entanglement distribution. The velocity of light ultimately limits communication. 

The third generation repeaters use deterministic entanglement distribution instead of the probabilistic ones. Considering the channel losses, the way to do deterministic entanglement distribution is to transmit encoded error correctable states between repeaters. Parity codes and codes based on cluster states can achieve this purpose. 

The repeater schemes are simple point to point network-based schemes, and they will not be sufficient to form a complex quantum network for real world applications. We may need a network model where both parties suspect a route between them but do not know it exactly. This may be dynamic because the relative costs and availabilities are liable to change. There is also the possibility that multiple paths are likely to exist between the concerned parties, and they could be attempted in a superposition fashion. This will increase both the capacity and robustness \cite{bib1}.

In 2015, Hensen group in the Netherlands demonstrated a loophole-free Bell test using electron spins separated up to 1.3km \cite{Hensen2016}. Their method using the Nitrogen vacancy (NV) centre could potentially implement device-independent quantum-secure communication as the same method can be handy in entanglement purification and distillation. Herald Weinfurter et al. in 2020 demonstrated entanglement generation between Rubidium (\textsubscript{87}Rb) atom and a photon at telecom wavelength (\textit{S }band- 1522nm) and transmission of the same up to 20km through an optical fibre \cite{PhysRevLett.124.010510}. In 2020, Jian Wei-Pan's group in China demonstrated the entanglement of two atomic ensembles that function as quantum memories storing quantum states via photon transmission over optical fibres. This was realised through field-deployed fibres via two-photon interference and over 22 kilometres and 50 kilometres for single photon interference. This experiment holds the promise of an atomic quantum network over longer distances \cite{Yu2020}.

In the document about the strategic vision of America's Quantum Networks, the US Government has stated their plan to enable quantum networks by 2025 \cite{USA}. The plan extends to realising quantum internet links to leverage networked quantum devices in another twenty years. At the same time, European Union has also laid down the plans for a European Quantum Communication Network in 2019 \cite{Riedel2019}. The agreement was formed between European Commission and European Space Agency to create a secure, pan-European quantum communication infrastructure that would comprise a series of quantum communication networks connecting institutions, critical infrastructures and sensitive communication and data sites in Europe. 

Quantum repeaters are expected to span several hundred kilometres forming metropolitan networks in the next five years. In another one or two decades, we may see quantum entanglement distribution over a thousand kilometres on the ground though quantum purification might not be guaranteed \cite{PhysRevLett.120.030503}. Quantum satellites and satellite-based entanglement distribution will play a key role in realising a global quantum internet \cite{Rabbie2022}. We will analyse them in the next section.

\section{Satellites}\label{sec8}

The network infrastructure for a global quantum internet is limited to the distance to which quantum information can be successfully transmitted. The inability to establish quantum repeaters spanning different geographies is a major limiting factor for the quantum internet. Losses in photonics of the optical fibres is another factor that restricts the quantum repeaters. The solution lies in quantum satellites which can work as space based quantum repeaters. A combination of terrestrial and space based segments is pivotal for the emanation of the quantum internet \cite{Sidhu2021, Belenchia2022}. 

The concept of using satellites for quantum communication is as old as mid 1990s. A research team led by Richard Hughes first suggested the possibility of using satellites for quantum key distribution \cite{879387}. European Union has conducted a lot of research in this area, ranging from Space QUEST project in 2004 to experiments onboard the International Space Station (ISS) \cite{Ursin2007, Ursin2009, Joshi2018, Josep}

Chinese Academy of Sciences launched QUESS (Quantum Experiments at Space Scale) in 2016. This mission used a low earth orbit satellite named Micius-I. The mission demonstrated a secure satellite-to-ground exchange of cryptographic keys between Micius-I and multiple ground stations in China, with a kHz key rate over 1200km, around 20 orders of magnitudes greater than that of an optical fibre channel. It also showed two-photon entanglement distribution to ground stations separated by 1200km and a violation of Bell inequality of $2.37 \pm 0.09$ under strict Einstein locality conditions. Quantum teleportation of single photon qubits also has been demonstrated through an uplink channel for ground-to-satellite teleportation for a distance up to 1400km \cite{Yin2017}.

For a global quantum internet, it will require a network of satellites. Large satellites like Micius are expensive to develop. An alternative proposal is to use small satellites called CubeSats, which are miniaturised nano satellites of dimensions 10cm $\times$ 10cm $\times$ 10cm cubic units and mass less than 1.5kg per unit. Centre for Quantum technologies has successfully demonstrated QKD using CubeSats \cite{Oi2017}. Further projects in this direction are announced as part of UK Singapore collaboration \cite{Bedington2016, Morong2012}. Other research projects in this area include QUBE, a German project to implement a downlink from satellite to earth for the exchange of encryption keys. Q\textsuperscript{3}Sat is an Austrian project to produce low-cost quantum keys, which uses only two single photon detectors and one active downlink every two orbits \cite{Neumann2018}. The attraction of this project is its low cost, which makes it affordable for starters. NanoBob mission is a French-Austrian collaboration that aims for a versatile QKD receiver in space \cite{Kerstel2018}. Canada is launching QUYSSat (Quantum Encryption and Science Satellite), a micro satellite to advance the science of quantum technologies for QKD and long-distance quantum entanglement. This is viewed as a starter to large scale missions and quantum networks \cite{Rideout2012}. Italian Space Agency and University of Padua Italy came up with a proposal to add reflectors and other simple equipment to regular satellites. A single photon exchange of onboard corner cube retroreflectors (CCR) has been demonstrated between low earth orbit and a ground station. This technique has been extended to demonstrate single photon transmission to middle earth orbit and even up to GNSS orbit. They have demonstrated this with lower error rates for quantum cryptography. Proof‐of‐principle demonstrations of satellite quantum communications have also been performed using other satellites apart from CCR \cite{PhysRevLett.115.040502}. 

NASA is one of the pioneers in satellite quantum communications. They pursue multiple projects targeting the development of a global quantum communication infrastructure. This involves using low earth orbit platforms like International Space Station (ISS) to distribute quantum entanglement between ground stations separated by thousands of kilometres. NASA also plans to use spacecraft in higher orbits to connect ground stations separated by longer distances. This will consist of a high clock‐rate source of high-fidelity entangled photon pairs, a detector array that will perform quantum state tomography on the photon pairs, and a dual‐telescope gimbal system. Telescopes will need exceptional pointing accuracy to overcome the high losses from high clock‐rate sources. It will also require high photon throughput on both the transmitter and receiver ends \cite{TURYSHEV2007}. NASA has also planned a Deep Space Quantum Link (DSQL) to do experiments to explore relativistic effects on quantum systems \cite{Mazzarella2021}. The project aims to access the regime where special and general relativity effects affect the outcome of quantum processes such as teleportation, entanglement distribution, and the violation of the Bell's inequalities. These effects will have a more profound experimental impact over long‐baseline channels between orbits. DSQL will also serve as the Lunar Gateway, a space station orbiting the moon proposed to establish a quantum link with Earth‐based ground stations or high‐altitude platforms orbiting the Earth \cite{Mazzarella2020}. This technology will require quantum communications systems with high-end performance. Lunar–Earth links will need a photon pair production rate of 100MHz to overcome the high loss and deliver more photon counts than background events. Performance of DSQL will be significantly benefitted using quantum memories. 

Drone based QKD is a recent trend in this area, which does not require clear weather conditions like satellite-based systems. Currently, entanglement distribution up to 1km has been experimentally demonstrated \cite{Xue:21}. They also have the potential to connect satellites with ground networks \cite{Liu2020}.  


\section{Integrated photonic platforms}\label{sec10}

While the elementary building blocks for quantum communication explained previously were decscribed in discrete terms as isolated buildling blocks, ultimately it will be necessary to combine them together into integrated platforms. In the same way that photonic quantum computing experiments are now largely shifting to integrated platforms owing to their inherent stability and miniaturisation, we expect the same to emerge in the context of quantum communications infrastructure, where the optical processing at communications nodes is tightly integrated.

Integrated waveguide architectures, typically etched into silicon wafers, are a leading candidate for this. These have become the platform of choice for optical quantum information processing, especially in the private sector. They have the advantage that they can piggyback off existing CMOS fabrication technology, allowing the infrastructure of an existing multi-trillion dollar economy to be utilised, rather than having to develop new manufacturing techniques from scratch.

Silicon based platforms are the dominant test beds for quantum technologies, inspite of the key challenges like photon losses.\cite{Wang2020} Other integrated platforms like Lithium Niobate has gained prominence in recent years due to development in thin film technology \cite{Zhang2019}. In addition to these there has been a surge of new integrated photonic platforms in recent years that include II-V Quantum dots, AlGaAsOI, Tantalum Pentoxide and Nitogen Vacancy centres in Diamond. We are still away from predicting a platform that will dominate the quantum computers like silicon which dominated the classical computers \cite{Moody2022}

While some aspects of quantum information processing are challenging to integrate in this manner -- most notably fast-feedforward and quantum memories -- no doubt new techniques for achieving these will be developed as a byproduct of the optical quantum computing economy into which companies such as PsiQuatnum and Xanadu are havily investing. As these developments in optical integration advance, they will find direct utility in the complimentary field of quantum communication, which has highly overlapping technological requirements.

\section{The future of the quantum internet}\label{sec11}

We currently stand at the very early stages in the development of quantum communications infrastructure, limited by current engineering technology. Most applications of quantum communication thus far have been in the form of quantum key distribution or other similarly simple quantum protocols. Although these demonstrations are what currently dominate the headlines, they are not the ultimate long-term goal of the quantum internet.

Classical computers only began to reach their full potential when we started networking them together, from which the internet emerged, enabling resource sharing, cloud computing, decentralisation of compute infrastructure, and massively improved efficiency and utility as a result. It is expected that quantum computers will similarly only realise their full potential when they too are interconnected by a global quantum internet.

But for quantum computers the payoff associated with networking them is far more attractive. Classical compute resources, whether networked or not, are given by the sum of the parts -- compute power is rouhgly linear in the number of transistors. But for quantum computers we know that in the best case scenarios compute power grows exponentially with the number of contributing qubits. That is, if quantum computers are networked together to act as one, their unified compute power is exponentially greater than the sum of the parts. This differential between exponential and linear scaling is in effect the price incentive associated with building the necessary infrastructure to interconnect them. Since this differential is an exponential one, we anticipate that in a future world with widespread quantum computing resources, the market incentive for building quantum communications infrastructure will be enormous, for which astronomical compute power is the payoff.

The simplest way to visualise how this might work is by considering the graph state model for quantum computing. Let each node in the network have some graph state, say a lattice. When networked using entanlgement links, these lattices can be fused together to create a far larger distributed lattice structure. On a rectangular lattice, the lattice's height corresponds to the number of computational qubits, while its width corresponds to circuit depth. That is by fusing graph topologies appropriately we can increase both the number of logical qubits in a computation and the effective circuit depth, which specifies algorithmic complexity.

In current classical cloud-based computing, security is largely based on trust. We must have confidence that the server holding or processing our data is not stealing or misusing it. However trust is a very weak form of security. Ideally, we would like stronger forms of security than simply trust. Homomorphic encryption is a technique by which encrypted data can be processed in encrypted form, without a server having to first decrypt it. Classically, while possible, this is prohibitive, given the resource overheads involved. However quantum mechanically there are proposals for quantum homomorphic encryption (QHE) that are information-theoretically secure, with highly efficient resource overheads \cite{bib:homo, bib:opt_homo, BJe15, DSS16}. That is, quantum future of cloud-based computing promised far stronger security than is possible even in principle on classical hardware.

Beyond this, blind quantum computing (BQC) is possible, whereby a client can securely specify both their data and the desired algorithm in encrypted form, without the server learning either \cite{bib:bqc}. Universal blind computing is known to not be possible using only classical resources.

While these may seem minor points, they have major implications for the development of the field. It implies that strategic competitors, say adversarial world powers, can safely license one anothers quantum hardware for outsourced computation with provable security. In the classical era this is something that competitors would not dream of doing using sensitive information.

This future is inevitably quite distant. The required technologies are largely in their infancy and highly limited in their applicability. But the motivation for pursuing and advancing these technologies is clear, since their outcome in the quantum era will be of far greater significance and with far greater return than what we have observed with the classical internet.

In quantum communications networks, Bell pairs -- two-qubit maximally entangled states -- are a universal resource for quantum communication, since they can be used to perform quantum state teleportation, thereby enabling the communication of any other state. Since Bell pairs are identical and a universal resource, it is expected they will constitute the fungible commodity of the distributed quantum economy. We expect their market value to be determined by the differential between what quantum resources are able to do independently versus what they are able to do in unison, a differential that we know is an exponential one. This will form the economic driving force of the quantum internet and it is clear its future value will be enormous.


\bibliography{sn-bibliography}


\end{document}